\documentclass{emulateapj}
\usepackage{natbib}
\usepackage{apjfonts}
\begin{document}

\title{The M87 Black Hole Mass from Gas-dynamical Models of Space
  Telescope Imaging Spectrograph Observations\footnotemark[1]}

\author{Jonelle L. Walsh\footnotemark[2], Aaron J.
  Barth\footnotemark[3], Luis C. Ho\footnotemark[4], Marc
  Sarzi\footnotemark[5]}

\footnotetext[1]{Based on observations made with the NASA/ESA
  \emph{Hubble Space Telescope}, obtained at the Space Telescope
  Science Institute, which is operated by the Association of
  Universities for Research in Astronomy, Inc., under NASA contract
  NAS 5-26555. These observations are associated with program
  GO-12162.}

\footnotetext[2]{Department of Astronomy, The University of Texas at
  Austin, 2515 Speedway, Stop C1400, Austin, TX 78712-1205, USA;
  jlwalsh@astro.as.utexas.edu}

\footnotetext[3]{Department of Physics and Astronomy, University of
  California at Irvine, 4129 Frederick Reines Hall, Irvine, CA
  92697-4574}

\footnotetext[4]{The Observatories of the Carnegie Institution for
  Science, 813 Santa Barbara Street, Pasadena, CA 91101, USA}

\footnotetext[5]{Centre for Astrophysics Research, University of
  Hertfordshire, AL10 9AB Hatfield, UK}

\begin{abstract}

  The supermassive black hole of M87 is one of the most massive black
  holes known and has been the subject of several stellar and
  gas-dynamical mass measurements; however the most recent revision to
  the stellar-dynamical black hole mass measurement is a factor of
  about two larger than the previous gas-dynamical
  determinations. Here, we apply comprehensive gas-dynamical models
  that include the propagation of emission-line profiles through the
  telescope and spectrograph optics to new Space Telescope Imaging
  Spectrograph observations from the \emph{Hubble Space
    Telescope}. Unlike the previous gas-dynamical studies of M87, we
  map out the complete kinematic structure of the emission-line disk
  within $\sim$40 pc from the nucleus, and find that a small amount of
  velocity dispersion internal to the gas disk is required to match
  the observed line widths. We examine a scenario in which the
  intrinsic velocity dispersion provides dynamical support to the
  disk, and determine that the inferred black hole mass increases by
  only 6\%. Incorporating this effect into the error budget, we
  ultimately measure a mass of $M_\mathrm{BH} = (3.5^{+0.9}_{-0.7})
  \times 10^9\ M_\odot$ (68\% confidence). Our gas-dynamical black
  hole mass continues to differ from the most recent stellar-dynamical
  mass by a factor of two, underscoring the need for carrying out more
  cross-checks between the two main black hole mass measurement
  methods.

\end{abstract}


\section{Introduction}
\label{sec:intro}

Supermassive black holes are thought to play a fundamental role in the
growth and evolution of galaxies \citep{Silk_Rees_1998,
  DiMatteo_2005}. This idea is supported by the empirical
relationships between the mass of the black hole, whose gravitational
influence is limited to a very small region at the center of the
galaxy, and the large-scale properties of the host galaxy, such as the
bulge stellar velocity dispersion ($\sigma_\star$; e.g.,
\citealt{Ferrarese_2000, Gebhardt_2000b, Tremaine_2002,
  Gultekin_2009}) and the luminosity ($L_\mathrm{bul}$; e.g.,
\citealt{Dressler_1989, Kormendy_Richstone_1995, Marconi_Hunt_2003,
  Sani_2011}). Other interpretations have also been put forth to
explain the black hole mass $-$ bulge correlations. For example, the
relations may simply be the result of random mergers without the need
for additional physical processes like active galactic nuclei (AGN)
feedback \citep{Peng_2007, Jahnke_Maccio_2011}. Our understanding of
the interplay between black holes and their host galaxies is highly
incomplete. Thus, accurately characterizing the shape and cosmic
scatter of the relations is crucial for distinguishing between the
various theoretical interpretations, as well as for determining the
black hole mass function and space density of supermassive black
holes.

The black hole mass $-$ host galaxy correlations are composed of 87
black hole mass ($M_\mathrm{BH}$) measurements, which are most often
made by modeling the motions of stars or nuclear gas disks
\citep{Kormendy_Ho_2013}. In spite of the impressive number of
observations, the local black hole mass census is still incomplete,
notably for low and high-mass black holes, and there remain major open
questions. At the high-mass end, the slope, intrinsic scatter, and
possibly the functional form of the $M_\mathrm{BH}$ $-$ host galaxy
relations are not well constrained \citep{Gultekin_2009,
  Graham_Scott_2013, McConnell_Ma_2013}. Properly quantifying the
scatter at the high-mass end of the $M_\mathrm{BH}$ correlations may
be especially helpful as non-causal interpretations of the scaling
relations predict a decrease in the scatter at higher black hole and
galaxy masses \citep{Jahnke_Maccio_2011}. Furthermore, the current
$M_\mathrm{BH} - \sigma_\star$ and $M_\mathrm{BH} - L_\mathrm{bul}$
relationships make strongly divergent predictions for the black hole
masses in the most luminous and the highest-dispersion galaxies
\citep{Lauer_2007, Bernardi_2007}, leading to questions about which of
the two correlations is more fundamental. At least one of the scaling
relations must be wrong at the high-mass end, but currently it is not
known which or how to reconcile the difference. In addition, there
have been very few consistency checks between the stellar and
gas-dynamical methods, therefore it is unclear how much of the
intrinsic scatter is due to inconsistencies between stellar and
gas-dynamical measurements, or if there a systematic difference
between the masses derived using the two methods.

Only direct comparisons between the stellar and gas-dynamical methods
within the same galaxy can address some of these questions, but
presently such tests have only been attempted on a very small number
of galaxies. Stellar-dynamical black hole masses have been made for
both IC 1459 and NGC 3379, but the ionized gas turned out to be highly
disturbed and a useful gas-dynamical $M_\mathrm{BH}$ measurement could
not be obtained \citep{VerdoesKleijn_2000, Gebhardt_2000a,
  Cappellari_2002b, Shapiro_2006, vandenBosch_deZeeuw_2010}. In NGC
4151, the gas-dynamical determination \citep{Hick_Malkan_2008} is
generally consistent with the stellar-dynamical mass, but the authors
label their stellar-dynamical result as tentative because they were
unable to find single best fitting model \citep{Onken_2007}. Also, the
gas-dynamical measurement for the black hole in M81
\citep{Devereux_2003} agrees with a previous, but preliminary,
stellar-dynamical determination \citep{Bower_2000}. This leaves six
galaxies (NGC 3227, NGC 3998, NGC 4258, NGC 4335, M87, Cen A) for
which the two types of black hole mass measurements can be
meaningfully compared. The stellar and gas $M_\mathrm{BH}$
measurements for NGC 3227, NGC 4258, and Cen A are consistent
\citep{Davies_2006, Pastorini_2007, Neumayer_2007, Hick_Malkan_2008,
  Siopis_2009, Cappellari_2009}, while the stellar-dynamical mass
exceeds the gas-dynamical determination by a factor of $\sim$2 $-$ 5
for the remaining three galaxies \citep{Macchetto_1997,
  VerdoesKleijn_2002, deFrancesco_2006, Gebhardt_2011, Walsh_2012}.

In particular, the supermassive black hole in giant elliptical galaxy
M87 has been the subject of numerous mass determinations, beginning
with the pioneering work of \cite{Sargent_1978} and \cite{Young_1978},
who suggested the presence of a $\sim$$(3 - 5) \times 10^9\ M_\odot$
black hole based on the surface brightness and stellar velocity
dispersion profiles. There were concerns associated with these and
other early stellar-dynamical black hole mass measurements (see review
by \citealt{Kormendy_Richstone_1995} and references therein), but a
much improved stellar-dynamical measurement was recently obtained by
\cite{Gebhardt_Thomas_2009} and \cite{Gebhardt_2011} using
three-integral, axisymmetric Schwarzschild models that also include
the mass distribution of the dark halo. More specifically,
\cite{Gebhardt_2011} applied orbit-based models to high angular
resolution spectroscopy obtained with Gemini's Near-Infrared Integral
Field Spectrometer assisted by adaptive optics and additional
large-scale data \citep{Emsellem_2004, Murphy_2011}, finding
$M_\mathrm{BH} = (6.6 \pm 0.4) \times 10^9\ M_\odot$.

In addition to the stellar-dynamical $M_\mathrm{BH}$ measurements,
there are also two gas-dynamical measurements for the black hole in
M87. In fact, M87 was the first gas-dynamical target with the
\emph{Hubble Space Telescope} (\emph{HST}). Using spectra acquired
with the Faint Object Spectrograph (FOS), \cite{Harms_1994} modeled
the emission-line kinematics as a rotating disk to infer a central
mass of $(2.4 \pm 0.7) \times 10^9\ M_\odot$. A few years later,
\cite{Macchetto_1997} used spectroscopy from the Faint Object Camera
(FOC) and rotating, thin-disk models that, for the first time,
incorporated the important effects of the propagation of light through
the telescope and spectrograph optics to determine $M_\mathrm{BH} =
(3.2 \pm 0.9) \times 10^9\ M_\odot$.

The work of \cite{Harms_1994} and \cite{Macchetto_1997} were important
milestones for \emph{HST} and the field of supermassive black hole
detection. However, neither study was able to fully map out the
kinematic structure of the gas disk. \cite{Harms_1994} relied on just
six spectra obtained with the FOS and \cite{Macchetto_1997} used FOC
long-slit observations at three, non-contiguous, parallel positions
that only covered a fraction of the disk. Also, the disk inclination
has been a source of uncertainty in the M87 gas-dynamical black hole
mass. \cite{Harms_1994} adopted a value of $i = 42^\circ \pm 5^\circ$,
which was measured by \cite{Ford_1994} from \emph{HST} H$\alpha$
narrowband images, while \cite{Macchetto_1997} allowed the inclination
to be a free parameter in their dynamical models, but found a wide
range of acceptable values with $47^\circ < i < 65^\circ$. Perhaps
more importantly, later investigations of circumnuclear gas disks
revealed the prevalence of an intrinsic velocity dispersion
\citep[e.g.,][]{vanderMarel_vandenBosch_1998, VerdoesKleijn_2000,
  DallaBonta_2009}. Although the origin of the intrinsic velocity
dispersion is not known, it may be dynamically important, contributing
pressure support to the disk. Models that treat such systems as
simple, thin, dynamically cold disks will underestimate
$M_\mathrm{BH}$ \citep{Barth_2001, Neumayer_2007, Walsh_2010}. Only by
carrying out detailed modeling of the emission-line profiles,
including the contributions to the line widths from rotational and
instrumental broadening, will it be possible to determine if there is
significant intrinsic velocity dispersion in the M87 disk and its
subsequent effect on the gas-dynamical $M_\mathrm{BH}$ measurement.

Given that M87 is located at the extreme high-mass end of the
$M_\mathrm{BH}$ $-$ host galaxy relations and is a very important test
case for a gas/stars comparison study, the factor of $\sim$2
discrepancy between the best current stellar and gas-dynamical masses
is troubling. Furthermore, the mass of the black hole in M87 is a
crucial parameter for making inferences about the size of the
innermost stable circular orbit of the black hole and the black spin
from Very Long Baseline Interferometry observations
\citep{Doeleman_2012}. In this paper, we revisit the gas-dynamical
black hole mass measurement. We use new \emph{HST} data acquired with
the Space Telescope Imaging Spectrograph (STIS) and carry out more
comprehensive gas-dynamical modeling than has previously been
attempted for this galaxy. We describe the observations in \S
\ref{sec:obs}, the emission-line kinematics in \S
\ref{sec:kinematics}, the modeling procedures in \S
\ref{sec:gasmodeling} and \ref{sec:apptom87}, the results and error
analysis in \S \ref{sec:results}, and conclude in \S
\ref{sec:discussion_conclusions}. Throughout the paper, we assume a
distance to M87 of $D = 17.9$ Mpc in order to allow for a direct
comparison to the stellar-dynamical $M_\mathrm{BH}$ measurement by
\cite{Gebhardt_2011}.

\section{\emph{HST} Observations and Data Reduction}
\label{sec:obs}

We obtained new \emph{HST} STIS observations of M87 on 2011 June 7.
The observations have a wavelength scale of $0.554$ \AA\ pixel$^{-1}$
and a spatial scale of $0$\farcs$0507$ pixel$^{-1}$. We used the G750M
grating, centered on 6581 \AA, to provide coverage of the H$\alpha$
spectral region, which includes the [\ion{O}{1}] $\lambda\lambda 6300,
6364$, H$\alpha$, [\ion{N}{2}] $\lambda\lambda 6548, 6583$, and
[\ion{S}{2}] $\lambda\lambda 6716, 6731$ emission lines. However, the
[\ion{O}{1}] and [\ion{S}{2}] lines were weak, except at locations
very close to the nucleus, and we focus solely on the H$\alpha$ and
[\ion{N}{2}] lines for the remainder of the paper. We used the E1
aperture position to place the nucleus near the readout end of the
CCD, to avoid charge transfer efficiency losses. The STIS
\texttt{52x0.1} slit was placed at five parallel positions without any
space between adjacent positions, and was oriented at a position angle
(PA) of $51^\circ$ [within $15^\circ$ of the nuclear gas disk's major
axis as determined by \cite{Macchetto_1997}]. Between $3 - 5$ dithered
exposures were acquired at each slit position to aid in the removal of
cosmic rays, leading to exposure times that ranged from $1521 - 2911$
s per slit position.

The data from each subexposure were reduced by trimming the overscan
region, subtracting the bias and dark files, and applying flat-field
corrections. After this initial processing, we used {\tt LA-COSMIC}
\citep{vanDokkum_2001} to remove cosmic rays and hot pixels. We
completed the reduction of each subexposure by wavelength and flux
calibrating the data and rectifying for geometric distortions. The
reductions were carried out using individual IRAF\footnote[6]{IRAF
  is distributed by the National Optical Astronomy Observatory, which
  is operated by the Association of Universities for Research in
  Astronomy under cooperative agreement with the National Science
  Foundation} tasks within the standard Space Telescope Science
Institute (STScI) pipeline, with the exception of the cosmic-ray
cleaning step. The reduced, geometrically rectified, subexposures were
aligned and combined using the IRAF tasks {\tt IMSHIFT} and {\tt
  IMCOMBINE} in order to produce the final two-dimensional (2D) STIS
image at each of the five slit positions.

\begin{figure}
\begin{center}
\epsscale{1.1}
\plotone{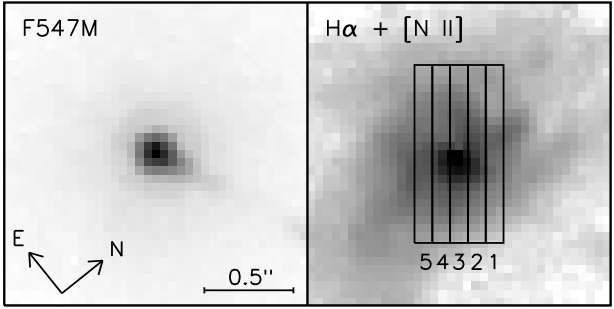}
\caption{\emph{HST} WFPC2/PC F547M (left) and continuum-subtracted
  H$\alpha$+[\ion{N}{2}] (right) images of the M87 nucleus. The images
  have been rotated such that the STIS instrumental $y$-axis points
  upward, and each box is 1\farcs7 on a side. The five STIS slit
  positions are overplotted on the continuum-subtracted image, and the
  length of the rectangles show the region over which the emission
  lines could be measured. \label{fig:wfpc2image}}
\end{center}
\end{figure}

Furthermore, from the \emph{HST} archive we retrieved Wide Field
Planetary Camera 2 (WFPC2) F658N, F547M, and F702W images, which were
originally made under programs GO-5122 and GO-5476 and had the M87
nucleus placed on the PC detector. The images in each filter were
taken as a sequence of two individual exposures with total exposure
times of 2700s, 800 s, and 280 s for the F658N, F547M, and F702W
images, respectively. The subexposures were combined and the cosmic
rays were rejected using the {\tt IMCOMBINE} and {\tt LA-COSMIC} IRAF
tasks. We then created a continuum-subtracted H$\alpha$+[\ion{N}{2}]
image by rotating the F658N and F702W images to a common orientation,
and subtracting a scaled version of the F702W image from the F658N
image. The scale factor was chosen such that a background region in
the continuum-subtracted image would have a mean flux as close to zero
as possible. In Figure \ref{fig:wfpc2image}, we present the
emission-line image with the location of the STIS slits overlaid, as
well as the F547M WFPC2 image. A compact central source and an
extended disk structure with several spiral-like wisps is seen in the
continuum-subtracted image, as was previously found by
\cite{Ford_1994}.

\section{Emission-line Kinematics}
\label{sec:kinematics}

From each of the 2D STIS images, we extracted spectra from individual
rows out to $\sim$0\farcs5 from the slit center, which is as far as
the emission lines were detectable. At locations where the emission
lines became weak, we binned together three STIS rows in order to
improve the signal-to-noise ratio (S/N). Prior to measuring the
emission-line kinematics, we subtracted the starlight continuum on a
row-by-row basis by fitting a line between wavelengths $6295 - 6862$
\AA\ excluding the regions near the emission lines, then subtracting
this fit from the spectrum. Due to the low S/N in the continuum and
the small wavelength range, a straight line is a sufficient
description of the continuum near H$\alpha$ and we do not attempt a
more complicated subtraction. Below we detail our spectral fitting
procedure and describe the observed velocity fields.

\subsection{Spectral Fitting Method}
\label{subsec:specfit}

We simultaneously fit three single Gaussians to the
continuum-subtracted H$\alpha$ and [\ion{N}{2}] emission lines using
the {\tt MPFIT} library in IDL \citep{Markwardt_2009}, which employs a
Levenberg-Marquardt least-squares minimization technique. We then
performed a Monte Carlo, adding random Gaussian noise to the spectrum
based upon the model residuals, and refit the new spectrum to
determine the velocity, velocity dispersion, and flux for each of the
three emission lines during each iteration. From the resulting
distributions, we were able to determine whether the [\ion{N}{2}]
$\lambda 6548$ and $\lambda 6583$ velocities were consistent within
2$\sigma$. If the velocities were consistent, we calculated the mean
and the 2$\sigma$ errors from the combined distribution of the
[\ion{N}{2}] velocities, which was taken to be the final velocity and
error for that spectral row. The same process was adopted for the
[\ion{N}{2}] velocity dispersions and for $3\times$ the flux of the
[\ion{N}{2}] $\lambda 6548$ line and the flux of the [\ion{N}{2}]
$\lambda 6583$ line. If instead the kinematics of the [\ion{N}{2}]
lines were not consistent, we fit progressively more constrained
models (such as tying together the [\ion{N}{2}] velocity widths) until
the [\ion{N}{2}] kinematics were consistent. With our approach, we are
assuming that the [\ion{N}{2}] velocities should be equal, the
[\ion{N}{2}] velocity dispersions should be the same, the fluxes of
the [\ion{N}{2}] lines should maintain a $3:1$ ratio (as dictated by
atomic physics), and that any differences that are seen are due to
fitting error. We chose not to fit Gaussian models that enforce these
constraints from the start because we found doing so produced very
small errors on the kinematics - typically we measured velocity errors
of only a few km s$^{-1}$ for the good S/N spectra (while the STIS
pixel scale is $\sim$25 km s$^{-1}$).

We also tested whether the spectral fits could be significantly
improved with the addition of a broad H$\alpha$ component described by
a single Gaussian. Including a broad component either caused the
$\chi^2$ per degree of freedom ($\chi^2_\nu$) to decrease by less than
15\%, produced a broad component with zero flux, or resulted in
unreasonably small fluxes and/or line widths for the H$\alpha$ and
[\ion{N}{2}] narrow components. We cannot fully rule out the existence
of a compact broad-line region, but the data do not require the
presence of a physically distinct broad-line component, similar to the
giant elliptical galaxy M84 \citep{Walsh_2010}. In both these
galaxies, the very broad central line widths are consistent with
coming from the inner portion of the rotating disk. Thus, the
kinematics presented here come from single Gaussian fits to the
H$\alpha$ and [\ion{N}{2}] lines only without the inclusion of a model
for a broad component.

\begin{figure}
\begin{center}
\epsscale{1.1}
\plotone{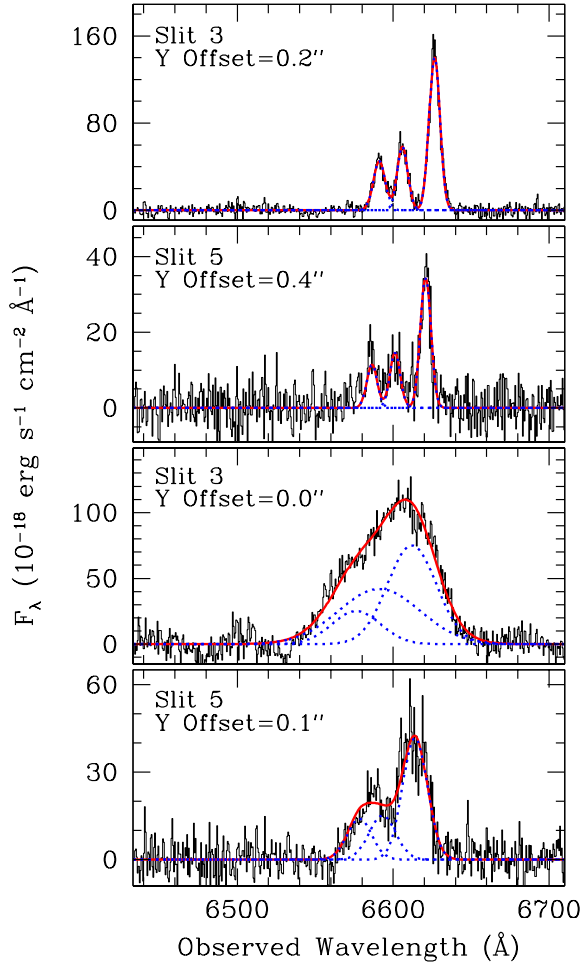}
\caption{Example fits to the H$\alpha$ and [\ion{N}{2}] emission
  lines. Blue dotted lines show the three single Gaussians that were
  fit, and the red solid line is the sum. The top two panels depict
  the range in quality of spectra from which kinematic measurements
  were made. Close to the nucleus and near the center of slit position
  5, it was difficult to accurately and uniquely decompose the spectra
  into into the H$\alpha$ and [\ion{N}{2}] components, as can be seen
  in the bottom two panels. Therefore, the kinematics measured from
  these central rows were excluded during the dynamical modeling.
  \label{fig:exspec}}
\end{center}
\end{figure}

While the above fitting method worked well for many of the observed
spectra, the spectra extracted near the center of slit positions 2, 3,
and 4 were complex with severely blended emission lines. In addition,
some of these rows showed artifacts, in the form of a periodic
fluctuation in the flux, from the geometric rectification step of the
data reduction \citep{Dressel_2007}. Moreover, spectra extracted near
the center of slit position 5 exhibited blended [\ion{N}{2}] $\lambda
6548$ and H$\alpha$ emission lines and a noisy [\ion{N}{2}] $\lambda
6583$ line that may be double-peaked. It was therefore difficult to
decompose the spectra into H$\alpha$ and [\ion{N}{2}] components with
accurate and unique mean velocities, velocity dispersions, and fluxes,
even when applying constraints to the three single Gaussians that were
being fit. Consequently, we consider the measurements from the central
rows of slits 2, 3, 4, and 5 to be unreliable and we do not use them
to constrain our gas-dynamical models. The innermost kinematic
measurements provide relatively little useful information when
constraining the disk models because the line profiles near the
nucleus are dominated by rotational broadening and blurring by the
point-spread function (PSF). For completeness, however, we incorporate
into our final error budget the effect on the black hole mass when the
measurements deemed as unreliable are used to constrain the dynamical
model as well. In Figure \ref{fig:exspec}, we show example fits to
spectra extracted from slit positions 3 and 5, including rows where
the kinematic measurements are considered unreliable.

\subsection{Observed Velocity Fields}
\label{subsec:obsvelfields}

From the spectral fitting, we were able to map out the emission-line
velocity, velocity dispersion, and flux as a function of location
along the slit for each of the five slit positions. The velocity
curves reveal that the gas participates in regular rotation, and the
gas to the northeast of the nucleus is redshifted relative to the
galaxy's systemic velocity while the gas located to the southwest is
blueshifted. There is a steep velocity gradient of $\sim$1200 km
s$^{-1}$ across the inner 0\farcs2 of slit position 3. The observed
velocity dispersions, which include contributions from rotational and
instrumental broadening, range from $\sim$150 -- 450 km s$^{-1}$,
although larger velocity dispersions are likely present as the line
widths continue to rise toward the nucleus. However, we were unable to
obtain reliable measurements of the central velocity dispersions
because the emission lines were severely blended together in the
nuclear spectra. In addition, the observed velocity dispersion profile
appears asymmetric, perhaps as a result of interactions with the
one-sided jet in M87. For example, the gas has velocity dispersions of
461 km s$^{-1}$ and 263 km s$^{-1}$ at locations $-$0\farcs1 and
0\farcs1 away from the nucleus in slit position 3.

\section{Gas-dynamical Modeling: Overview}
\label{sec:gasmodeling}

The observed velocity fields are modeled assuming that the gas
participates in circular rotation in a thin disk. We provide a brief
overview of the modeling here, but refer the reader to
\cite{Barth_2001} and \cite{Walsh_2010} for further details. At each
radius ($r$) in the disk, the circular velocity is calculated relative
to the galaxy's systemic velocity ($\upsilon_\mathrm{sys}$) based on
the enclosed mass, which depends on the black hole mass, the stellar
mass profile, and the mass-to-light ratio ($\Upsilon$). The model
velocity field is created on a pixel grid that is oversampled relative
to the size of a STIS pixel, and is projected onto the plane of the
sky for a given value of the disk inclination angle ($i$).

The intrinsic line-of-sight velocity profiles are assumed to be
Gaussian before passing through the telescope optics and are generated
on a velocity grid with a bin size that matches the STIS scale. The
intrinsic line profiles are centered on the projected line-of-sight
velocity at each point on the model grid and weighted by the
emission-line flux distribution. In addition, the widths of the
intrinsic line profiles are set based upon contributions from the
thermal velocity dispersion of the gas and the velocity dispersion due
to turbulent motion internal to the gas disk.

Next, the model velocity field is ``observed'' in a manner that
matches the STIS observations. This synthetic observation includes
accounting for the blurring by the telescope PSF by convolving a Tiny
Tim \citep{Krist_Hook_2004} PSF model with each velocity slice of the
line profile grid. Also, we propagate the line profiles through the
finite width of the STIS slit by applying shifts to the velocity
depending on the location along the slit width that the photon enters
\citep{Maciejewski_Binney_2001}. After rebinning to the STIS pixel
size, we are left with a model 2D spectral image similar to the
observed STIS data. At this stage, we convolve the model 2D STIS image
with the Tiny Tim CCD charge diffusion kernel in order to account for
the charge that is spread between non-subsampled, neighboring pixels.

Finally, we extract spectra on a row-by-row basis from the model 2D
image and fit a single Gaussian to the emission line, analogous to the
measurements that were made from the data. We thus are able to measure
the model velocity, velocity dispersion, and flux as a function of
position along the slit. The best model is taken to be the one that
most closely matches the observations in the $\chi^2$ sense. We use
the downhill simplex algorithm by \cite{Press_1992} to minimize
$\chi^2$ and determine the best-fit parameters. We separately optimize
the fits to the observed line widths and fluxes before determining the
black hole mass from the model fit to the observed radial
velocities. When fitting to the observed velocities, the model
parameters are $M_\mathrm{BH}$, $\Upsilon$, $i$,
$\upsilon_\mathrm{sys}$, $\theta$, $x_\mathrm{offset}$, and
$y_\mathrm{offset}$. Here, $\theta$ is the angle between the slit and
the projected major axis of the disk, and $x_\mathrm{offset}$ and
$y_\mathrm{offset}$ are the offsets between the center slit and the
nucleus.

\section{Gas-dynamical Modeling: Application to M87}
\label{sec:apptom87}

Gas-dynamical models for M87 were calculated on a pixel grid
oversampled by a factor of $s=6$ relative to the STIS pixel size. We
found that preliminary models with a subsampling factor of $s=6$ were
able to best reproduce the observed velocities and emission-line
fluxes, and could be run in a reasonable amount of time. Thus, we
selected $s=6$ for the final models. We also incorporated the effect
on $M_\mathrm{BH}$ of our subsampling factor choice into the error
budget, which will be summarized in \S \ref{subsec:mbherror}. In
addition, we used a PSF generated with Tiny Tim for a monochromic
filter passband at 6600 \AA. The model is an $s=6$,
0\farcs4$\times$0\farcs4 portion of the full STIS PSF. Such a model
excludes part of the extended PSF wings, with 12\% of the total flux
being omitted. As will be shown in \S \ref{subsec:mbherror}, the black
hole mass is not strongly dependent on the size of the PSF model, but
because the computation time increases rapidly with increasing PSF
size, we chose to use a 0\farcs4-diameter PSF. Finally, we fit disk
models over the entire radial range covered by the observed
kinematics. With a fitting radius of 0\farcs5, not only is there an
adequate number of measurements to constrain the dynamical models, but
also the disk model continues to provide a good description of the
observations out to this radius. As discussed previously, we exclude
the uncertain kinematics measured from the central rows during the
modeling.

Below we provide additional details about the M87 gas-dynamical
models, including the stellar contribution to the circular velocities,
the parameterization of the intrinsic emission-line flux distribution,
the components that make up the width of the line profiles, and our
approach for allowing the turbulent velocity dispersion to be
dynamically important.

\subsection{Stellar Mass Model}
\label{subsec:stellarmass}

We determined the stellar contribution to the gravitational potential
using the $V$-band surface brightness profile of M87 from
\cite{Kormendy_2009}. The profile combines measurements from
ground-based images with high-resolution \emph{HST} WFPC1/PC data,
which were originally presented in \cite{Lauer_1992} and have already
been deconvolved by the telescope PSF.  In addition to the surface
brightness profile, \cite{Kormendy_2009} present the ellipticity
($\epsilon$) of the isophotes as a function of major axis radius. At
radii within 180\arcsec, the isophotes are nearly circular, with
$\epsilon < 0.2$. Therefore, we deprojected the surface brightness
profile assuming that the intrinsic density distribution of the
central regions could be modeled as the sum of spherically symmetric
components with Gaussian profiles. A similar approach has been used to
derive the luminous density distribution for past gas-dynamical black
hole mass measurements \citep[e.g.,][]{Sarzi_2001, Barth_2001,
  Coccato_2006, Walsh_2010}. To carry out the deprojection, we used
the Multi-Gaussian Expansion (MGE) software developed by
\cite{Cappellari_2002a}, which is based on the methods described in
\cite{Emsellem_1994}. In addition, the M87 nucleus is classified as a
Type 2 Low Ionization Nuclear Emission-line Region \citep{Ho_1997}. We
attribute the innermost compact Gaussian component of the MGE model,
with a dispersion of 3.5 pc, to light from the AGN, and exclude this
component from the stellar mass distribution.

\begin{figure}
\begin{center}
\epsscale{1.1}
\plotone{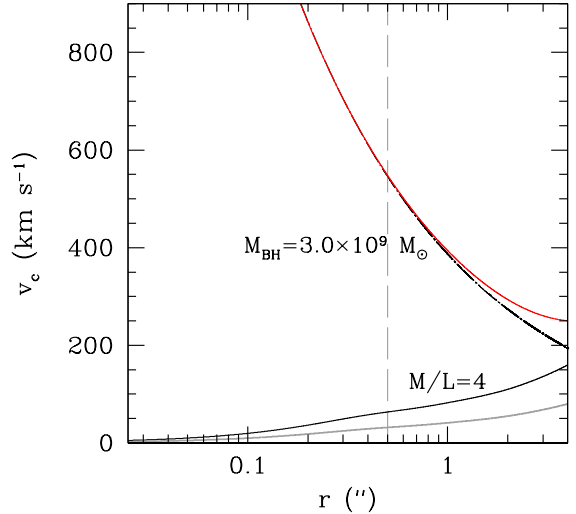}
\caption{The circular velocity as a function of radius due to the
  presence of a $3.0\times10^9\ M_\odot$ black hole (dot-dashed black
  curve) and the stars with $\Upsilon = 4$ (solid black line) and
  $\Upsilon = 1$ (grey solid line). The solid red curve shows the sum
  of the circular velocities from the black hole and the stars when
  assuming $\Upsilon = 4$ ($V$-band, solar). The dashed vertical grey
  line indicates the largest radius probed by the STIS data. Over this
  range, the black hole dominates the gravitational potential and the
  stellar contribution is negligible. \label{fig:stellarvc}}
\end{center}
\end{figure}

The final product of the multi-Gaussian deprojection of the surface
brightness profile is the stellar contribution to the circular
velocity in the galaxy, which is shown in Figure
\ref{fig:stellarvc}. The figure also provides the circular velocities
due to the presence of a $3.0\times10^9\ M_\odot$ black hole, which is
a plausible lower limit on the black hole mass from our
analysis. Clearly, the black hole's sphere of influence is well
resolved by our STIS observations, which is further supported by the
nearly Keplerian shape of the multi-slit velocity curves. In other
words, the black hole dominates the galaxy's gravitational potential
and the stellar contribution is negligible over the radial range
probed by the STIS kinematics. Consequently, we found that our
dynamical models are insensitive to the stellar mass-to-light ratio,
and we chose to fix $\Upsilon = 4$ ($V$-band, solar units) during the
model fitting. This value was selected using the $B-V$ color for the
galaxy from the Third Reference Catalogue of Bright Galaxies
\citep{RC3_catalog} and the relations presented in
\cite{Bell_2003}. Such a stellar mass-to-light ratio is typically
found at the centers of giant elliptical galaxies
\citep{Cappellari_2006}. We also tested the effect on $M_\mathrm{BH}$
when another fixed value for $\Upsilon$ is adopted, as well as when
$\Upsilon$ is allowed to be a free parameter. The results will be
described in \S \ref{subsec:mbherror}.

\subsection{Emission-Line Flux}
\label{subsec:flux}

The line-of-sight velocity profiles are weighted by the emission-line
flux distribution at each point in the disk. Commonly, the flux
distribution is described by analytic functions, or by creating a
continuum-subtracted H$\alpha$ $+$ [\ion{N}{2}] image that has been
deconvolved by the telescope PSF \citep[e.g.,][]{Barth_2001,
  Pastorini_2007, Marconi_2006, Wold_2006}. We experimented with
directly folding a deconvolved continuum-subtracted image constructed
from WFPC2/PC F702W and F658N images into the modeling, as well as
using ten analytic forms composed of two and three Gaussian [$F(r) =
F_0 \exp(-r^2 / 2r_s^2)$] and exponential [$F(r) = F_0 \exp(-r /
r_s)$] components. We used analytic models that represented
intrinsically circularly symmetric disks, which produced concentric
elliptical isophotes with constant position angle and axis ratio, and
more complicated models in which the isophotes of the individual
components were allowed to have different centers, position angles,
and axis ratios. The parameters associated with each of the analytic
functions were determined by calculating disk models following the
methods described in \S \ref{sec:gasmodeling}, then optimizing the fit
to minimize the $\chi^2$ computed from the comparison of the model and
observed emission-line fluxes.

We found that using a deconvolved continuum-subtracted image and
analytic models that represented intrinsically circularly symmetric
disks returned unacceptable fits to the observed emission-line flux
and velocities, and we do not consider these descriptions
further. While deconvolved H$\alpha$ $+$ [\ion{N}{2}] images have been
folded into gas-dynamical models successfully in the past
\citep[e.g.,][]{Barth_2001, DallaBonta_2009}, we experienced
difficulties during the deconvolution process, which was performed
using the IRAF task {\tt LUCY}. The deconvolution produced an artifact
in the form of a hole near the nucleus and subsequently the model
fluxes were smaller than the observed fluxes for a portion of the
disk. The remaining five analytic models, however, were able to match
the observed flux distribution, and we searched for the simplest model
that would simultaneously produce model velocities closest to the
observed radial velocities. Ultimately, we determined that the best
parameterization of the intrinsic emission-line flux was the sum of
two Gaussian components. In Table \ref{tab:fluxmodel}, we present the
amplitudes ($F_0$), scale radii ($r_s$) along the major axis, position
angles, axis ratios ($b/a$), and centroid positions for each of the
two components. We also include the effect of using the other four
analytic flux models in the final error budget for the black hole mass
in \S \ref{subsec:mbherror}.

\begin{deluxetable}{lcccccc}
\tabletypesize{\footnotesize}
\tablewidth{235pt}
\tablecaption{Emission-Line Flux Parameters \label{tab:fluxmodel}}
\tablehead{
\colhead{Component} & 
\colhead{$F_0$} &
\colhead{$r_s$} &
\colhead{$x_\mathrm{cen}$} &
\colhead{$y_\mathrm{cen}$} &
\colhead{PA} &
\colhead{$b/a$} \\
\colhead{} &
\colhead{} &
\colhead{(pc)} &
\colhead{(\arcsec)} &
\colhead{(\arcsec)} &
\colhead{($^\circ$)} &
\colhead{}
}

\startdata

Gaussian  &  26.5  &  6.6  &  0.05  &  $-0.02$  &  14  &  0.13  \\
Gaussian  &  1.0  &  23.7  &  0.00  &  0.02  &  191  &  0.77

\enddata

\tablecomments{The amplitude, $F_0$, is in arbitrary units. The center
  of the ellipse is described by $x_\mathrm{cen}$ and $y_\mathrm{cen}$
  relative to the location of the black hole. The position angle, PA,
  is in units of degrees clockwise, with PA $ = 0$ pointing along the
  length of the STIS slit.}

\end{deluxetable}

\subsection{Line Widths}
\label{subsec:linewidths}

We initially ran models in which the widths of the emission-line
profiles originate from rotational, instrumental, and thermal
broadening. The rotational and instrumental broadening effects are
explicitly included in our dynamical models when the line profiles are
propagated through the telescope and instrument. In particular,
rotational broadening occurs because light from different portions of
the disk get blended together within the slit, while instrumental
broadening occurs because of PSF effects, the charge diffusion between
neighboring pixels, and the velocity shifts that result from the
finite slit width. Prior to propagating the line profiles through the
telescope optics, we also include a thermal contribution to the
velocity dispersion of $\sigma_\mathrm{th} = 10$ km s$^{-1}$, which is
expected for gas with a temperature of $\approx$$10^4$ K.

In some parts of the disk, the predictions from these initial models
were smaller than the observed velocity dispersions. Similar results
have been found when constructing gas-dynamical models of other
galaxies as well \citep[e.g.,][]{vanderMarel_vandenBosch_1998,
  Barth_2001, VerdoesKleijn_2002, Tadhunter_2003, DallaBonta_2009},
although in some instances the rotational and instrumental broadening
alone is enough to explain the observed line widths (e.g.,
\citealt{Marconi_2001, Capetti_2005, Atkinson_2005, deFrancesco_2006},
2008). Therefore, we added a projected intrinsic velocity dispersion
($\sigma_\mathrm{p}$) in quadrature to $\sigma_\mathrm{th}$ before
propagating the line profiles through the telescope and
spectrograph. The projected intrinsic velocity dispersion was
determined by constructing disk models as discussed in \S
\ref{sec:gasmodeling} and fitting to the observed [\ion{N}{2}] line
widths. With our early model runs, we tried characterizing
$\sigma_\mathrm{p}$ as a constant $+$ exponential of the form
$\sigma_0 + \sigma_1 \exp(-r/r_0)$, which has often worked well for
other galaxies, but found that best-fit value for $\sigma_1$ was
$\sim$0 km s$^{-1}$. Thus, in the final model, the intrinsic velocity
dispersion was parameterized as a constant over the disk, with
$\sigma_\mathrm{p} = 170$ km s$^{-1}$.

\subsection{Asymmetric Drift Correction}
\label{subsec:asymmdrift}

Currently, there is no clear consensus for how to interpret the
physical nature of the intrinsic velocity dispersion, hence we explore
two scenarios that should bracket the range of possible masses for the
black hole in M87. In one case, we assume that the intrinsic velocity
dispersion does not affect the circular velocity, which could result
if the intrinsic velocity dispersion is due to local microturbulence
but the bulk motion of the gas remains in circular motion
\citep{vanderMarel_vandenBosch_1998}. In the second case, we assume
that the internal velocity dispersion is due to turbulent motion that
supplies pressure support to the disk, causing the observed mean
rotational speed ($\upsilon_\phi$) to be smaller than the local
circular velocity ($\upsilon_c$) for a given black hole mass
\citep[e.g.,][]{Coccato_2006, Neumayer_2007, Walsh_2010}. For M87, we
estimate an upper bound to the black hole mass by applying an
asymmetric drift correction following \cite{Barth_2001}. Under the
assumption that the gas motions are isotropic in the $r$ and $z$
(cylindrical) coordinates, the correction is

\begin{equation}
\label{eq:asymmdriftcorr}
\upsilon_c^2\ -\ \upsilon_\phi^2\ =\ \sigma_r^2 \bigg[ -\frac{d \ln \nu}{d \ln r}\ -\ 
\frac{d \ln \sigma_r^2}{d \ln r}\ -\ 
\bigg( 1\ -\ \frac{\sigma_\phi^2}{\sigma_r^2} \bigg) \bigg] \;.
\end{equation}

We take the intrinsic emission-line flux distribution to be a proxy
for the number density of tracer particles, or clouds in the gas disk,
$\nu$. Since the asymmetric drift correction depends on an
azimuthally-averaged value for the radial derivative of $\nu$, we
refit the observed fluxes with an analytic function composed of two
Gaussians (similar to the flux model presented in \S
\ref{subsec:flux}), but required that the isophotes have the same
center, position angle, and axis ratio. For a given intrinsic radial
velocity dispersion, $\sigma_r$, we computed $\sigma_\phi$ and the
projected velocity dispersion as outlined by \cite{Barth_2001}, before
adding the projected velocity dispersion in quadrature to
$\sigma_\mathrm{th}$ and propagating the line profiles through the
telescope and instrument. Although we initially attempted to
parameterize $\sigma_r$ as a constant $+$ exponential function, we
found that the intrinsic radial velocity dispersion was best described
as constant. By calculating disk models with an asymmetric drift
correction, and fitting to the observed line widths, we found a
best-fit value of $\sigma_r = 191$ km s$^{-1}$.

\newpage

\section{Modeling Results}
\label{sec:results}

After optimizing the model fits to the observed emission-line fluxes
and velocity dispersions, we fit disk models to the observed radial
velocities. The results returned by the $\chi^2$-minimization routine
seemed to strongly depend on the initial guesses for $M_\mathrm{BH}$
and $i$ in particular. Therefore, to ensure a global minimum was
found, we used a two-step process. First, we ran a coarse grid of
models with fixed $M_\mathrm{BH}$ and $i$ values while the other
parameters ($\theta$, $\upsilon_\mathrm{sys}$, $x_\mathrm{offset}$,
$y_\mathrm{offset}$) were allowed to vary. Second, we used the
parameter values of the top five models (with the lowest $\chi^2$)
from the grid search as the starting points to do five runs in which
all the parameters were allowed to vary. By doing these five runs, we
were able to distinguish between local and global minima.

\begin{figure*}
\begin{center}
\epsscale{1.0}
\plotone{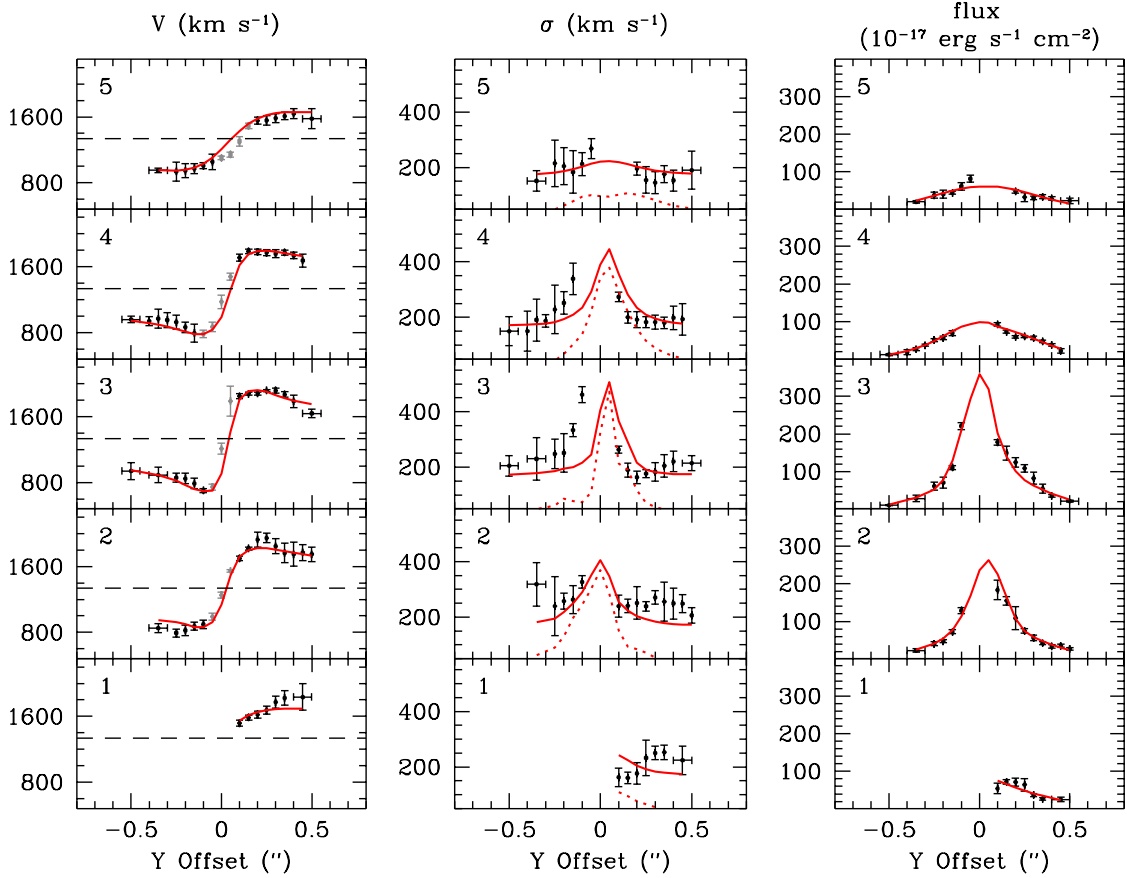}
\caption{The emission-line velocity, velocity dispersion, and flux
  (black points) measured as a function of location along the slit for
  the five slit positions. The numbers at the top left of each plot
  correspond to the slit locations shown in Figure
  \ref{fig:wfpc2image}, while the negative and positive Y Offsets
  correspond to the bottom and top half of the slits, respectively,
  depicted in Figure \ref{fig:wfpc2image}. We defined Y Offset $= 0$
  to be the spectral row with the largest continuum flux. Grey points
  indicate the velocity measurements that were considered unreliable
  and are plotted for reference, while the corresponding velocity
  dispersion and flux measurements are left off the figure. Spectral
  rows that were binned together to improve the S/N are shown with
  error bars along the $x$-axis. The spectra from the bottom portion
  of slit 1 had very poor S/N and we were unable to obtain kinematic
  measurements. The best-fit dynamical model for M87 without an
  asymmetric drift correction is overplotted with the red solid line,
  and the best-fit systematic velocity is displayed with the dashed
  black line. A single scaling factor has been applied to the model
  emission-line fluxes such that the median values of the model and
  data would match. The dotted red line shows the model line widths
  without the inclusion of an intrinsic velocity
  dispersion. \label{fig:datamodel}}
\end{center}
\end{figure*}

\begin{figure*}
\begin{center}
\epsscale{0.6}
\plotone{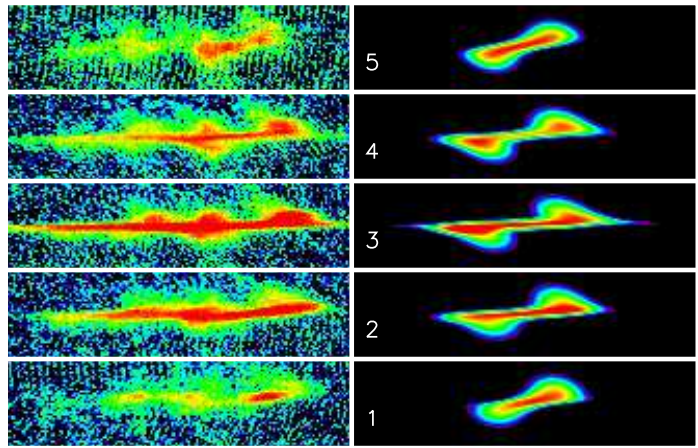}
\caption{Continuum-subtracted H$\alpha$ $+$ [\ion{N}{2}] region of the
  STIS data (left) and the synthetic 2D spectrum of the [\ion{N}{2}]
  $\lambda 6583$ emission line from the best-fit dynamical model
  without an asymmetric drift correction (right) at each of the five
  slit positions. Each box is 93 \AA\ along the dispersion
  (horizontal) direction and 2\farcs2 along the spatial (vertical)
  direction. The numbers correspond to the slit positions in Figure
  \ref{fig:wfpc2image}. \label{fig:datamodel_2d}}
\end{center}
\end{figure*}

The results of the best-fit model without an asymmetric drift
correction are summarized in Table \ref{tab:bestfitmodel} and the
model predictions are shown alongside the data in Figures
\ref{fig:datamodel} and \ref{fig:datamodel_2d}. We measure a black
hole mass of $M_\mathrm{BH} = 3.5\times10^9\ M_\odot$. The other model
parameters, including the disk inclination angle, are well
constrained. During the numerous preliminary model runs, the
parameters varied at most by 7$^\circ$, 9 km s$^{-1}$, 2$^\circ$,
0\farcs01, and 0\farcs04 for $i$, $\upsilon_\mathrm{sys}$, $\theta$,
$x_\mathrm{offset}$, and $y_\mathrm{offset}$, respectively. The model
has $\chi^2 = 66$, found by comparing the model predictions to the
observed velocities. Given that 62 velocity measurements were used to
constrain 6 model parameters, the $\chi^2$ per degree of freedom is
$\chi^2_\nu = 1.2$. As can be seen in Figure \ref{fig:datamodel}, the
model is able to reproduce the shape of the observed velocity curves
and the emission-line flux profiles very well. At some locations in
the disk, an intrinsic velocity dispersion is needed to match the
data, although this velocity dispersion is relatively small given the
large velocities found in M87. Moreover, our model for the intrinsic
velocity dispersion still severely underestimates the velocity
dispersions located $-$0\farcs10 and $-$0\farcs15 away from the center
of slits 3 and 4. We will examine this discrepancy and the possible
impact on the inferred black hole mass in \S \ref{subsec:mbherror}.

\begin{deluxetable}{cccccc}
\tabletypesize{\footnotesize}
\tablewidth{235pt}
\tablecaption{Disk Model Parameters \label{tab:bestfitmodel}}
\tablehead{
\colhead{$M_\mathrm{BH}$} & 
\colhead{$i$} &
\colhead{$\upsilon_\mathrm{sys}$} &
\colhead{$\theta$} &
\colhead{$x_\mathrm{offset}$} &
\colhead{$y_\mathrm{offset}$} \\
\colhead{$(M_\odot)$} & 
\colhead{($^\circ$)} &
\colhead{(km s$^{-1}$)} &
\colhead{($^\circ$)} &
\colhead{(\arcsec)} &
\colhead{(\arcsec)}
}

\startdata

$(3.5_{-0.7}^{+0.9}) \times 10^9$ & 42$_{-7}^{+5}$ &
1335$_{-9}^{+3}$ & 6$\pm2$ & $-0.02 \pm 0.01$ & $0.03_{-0.04}^{+0.02}$

\enddata

\tablecomments{These results give the best-fitting parameters for
  models calculated without an asymmetric drift correction. The errors
  given for the black hole mass are 1$\sigma$ uncertainties, and the
  errors provided for the other parameters correspond to the range of
  values found from all the preliminary models that were run. During
  the modeling, $\Upsilon$ was frozen at 4 ($V$-band solar units). A
  relative angle of $\theta = 6^\circ$ corresponds to a disk major
  axis position angle 45$^\circ$ east of north.}

\end{deluxetable}

\subsection{Uncertainty in the Black Hole Mass}
\label{subsec:mbherror}

\begin{figure}
\begin{center}
\epsscale{1.1}
\plotone{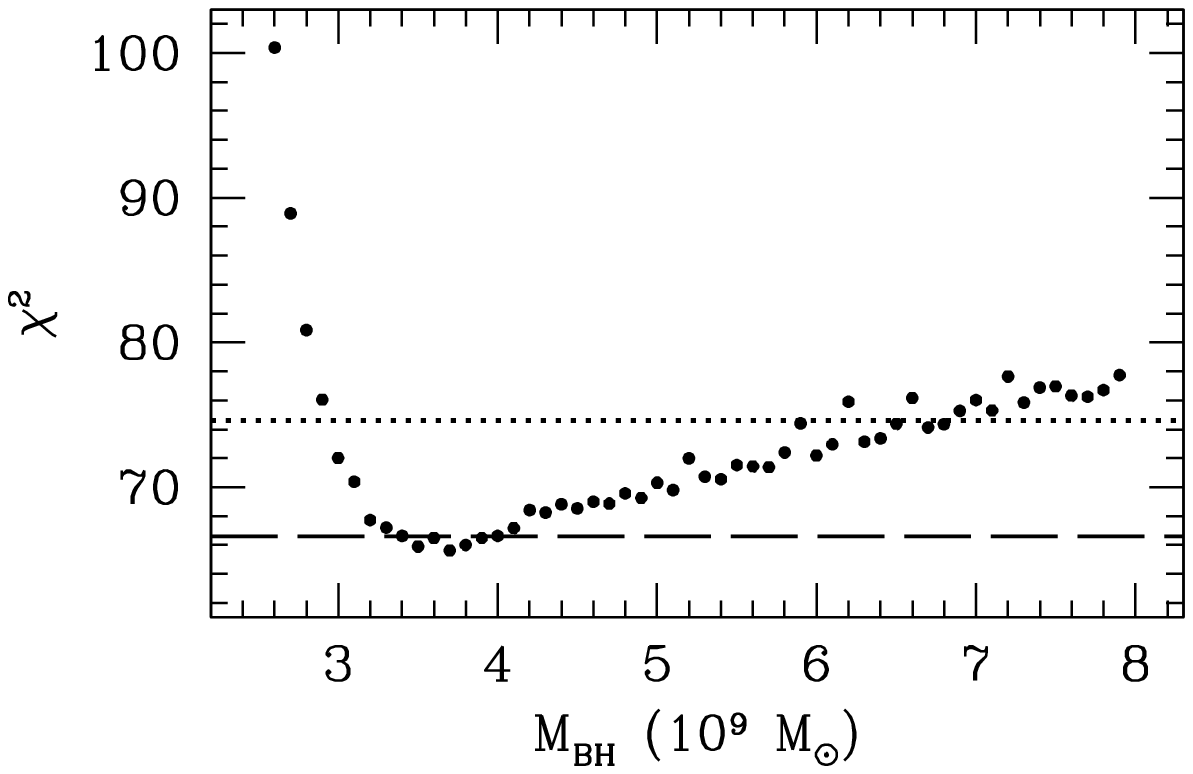}
\caption{Results of the disk models run without an asymmetric drift
  correction by fixing $M_\mathrm{BH}$ while allowing the other
  parameters to vary such that $\chi^2$ was minimized. The dashed and
  dotted lines denote where the minimum $\chi^2$ has increased by 1.0
  and 9.0, respectively, which corresponds to the 1$\sigma$ and
  3$\sigma$ uncertainties for one degree of
  freedom. \label{fig:chisq_mbh}}
\end{center}
\end{figure}

We measured the statistical errors on $M_\mathrm{BH}$ by holding
$M_\mathrm{BH}$ fixed while allowing the other parameters to vary such
that $\chi^2$ was minimized. During this process, we froze
$M_\mathrm{BH}$ to values between $(1.0 - 9.7) \times 10^9\ M_\odot$
in steps of $0.1 \times 10^9\ M_\odot$. We searched for the range of
$M_\mathrm{BH}$ values that caused the minimum $\chi^2$ to increase by
1.0 and 9.0 to estimate the 1$\sigma$ and 3$\sigma$ model fitting
uncertainties. Figure \ref{fig:chisq_mbh} presents the results of
these disk models, specifically the $\chi^2$ variation with
$M_\mathrm{BH}$ after marginalizing over the other
parameters. Ultimately, we determine that the range of black hole
masses is $(3.4 - 4.0) \times 10^9\ M_\odot$ (1$\sigma$ uncertainties)
and $(3.0 - 6.8) \times 10^9\ M_\odot$ (3$\sigma$ uncertainties).

These $M_\mathrm{BH}$ errors account for the random noise in the gas
dynamical models and the uncertainties associated with the parameters
$i$, $\upsilon_\mathrm{sys}$, $\theta$, $x_\mathrm{offset}$, and
$y_\mathrm{offset}$. However, there are additional sources of
uncertainty that are not incorporated into the statistical
errors. Below, we explore these sources of uncertainty and their
impact on $M_\mathrm{BH}$, in order to arrive at the final range of
possible black hole masses for M87.

\smallskip

\noindent \emph{Subsampling Factor}: We ran gas-dynamical models on
pixel grids that were subsampled by a factor of $s=2 - 10$, in
intervals of $s=2$. We found that the black hole mass varied between
$(3.5 - 3.9) \times 10^9\ M_\odot$ with a root-mean-square (rms)
scatter of $2.3 \times 10^8\ M_\odot$, which is 7\% of the best-fit
black hole mass.

\smallskip

\noindent \emph{PSF Size}: When calculating dynamical models using a
Tiny Tim PSF with diameters of 0\farcs3, 0\farcs4, 0\farcs5, and
0\farcs6, the rms scatter in the black hole mass was $1.2 \times 10^8\
M_\odot$. The PSF size has a minimal impact on the inferred black hole
mass, as the scatter is just 3\% of the best-fit mass.

\smallskip

\noindent \emph{Stellar Mass-to-Light Ratio}: The stellar contribution
to the galaxy's gravitational potential is negligible over the region
probed by the STIS data ($\sim$40 pc). As a result, our models were
unable to constrain the stellar mass-to-light ratio, and preliminary
model runs returned a wide range of values for $\Upsilon$. We chose to
fix $\Upsilon = 4$ ($V$-band, solar) during the final model run, but
also tested freezing $\Upsilon$ to 9, which is the stellar
mass-to-light ratio found for M87 by \cite{Murphy_2011} and adopted by
\cite{Gebhardt_2011}, and allowing the parameter to float. The test
produced in an rms scatter in the black hole mass of $1.3 \times 10^8\
M_\odot$, or 4\% of the best-fit $M_\mathrm{BH}$.

\smallskip

\noindent \emph{Emision-line Flux Model}: The intrinsic emission-line
flux distribution was modeled as the sum of two Gaussian components,
whose isophotes had different centers, position angles, and axis
ratios. An additional four analytic forms of varying complexity were
also able to reproduce the observed emission-line fluxes: two
exponentials, one Gaussian $+$ one exponential, three Gaussians, three
exponentials, and two Gaussians $+$ one exponential. These analytic
functions also represented elliptical isophotes with different
centers, position angles, and axis ratios. Although the quality of the
fit to the observed velocities varied significantly depending on the
intrinsic flux distribution used ($\chi^2$ ranged from 66 - 130),
there was a small effect on the black hole mass. The disk models with
different parameterizations of the emission-line flux distribution
returned best-fit masses between $3.1 \times 10^9\ M_\odot$ and $4.2
\times 10^9\ M_\odot$, with an an rms scatter of $5.0 \times 10^8\
M_\odot$, or 7\% of the best-fit black hole mass.

\smallskip

\noindent \emph{Differences Between the [\ion{N}{2}] and H$\alpha$
  Kinematics}: Gas-dynamical models were run using the [\ion{N}{2}]
kinematics because [\ion{N}{2}] $\lambda 6583$ was the strongest
emission line in our spectra. If gas-dynamical models are instead fit
to the H$\alpha$ kinematics, the black hole mass increases by 11\% to
$3.9\times10^9\ M_\odot$.

\smallskip

\noindent \emph{Including all Kinematic Measurements}: We had
difficulty fitting the spectra extracted near the center of slits $2 -
5$. We consider the kinematics and emission-line fluxes measured from
these rows unreliable and did not use them to constrain our final disk
model. When these velocities, velocity dispersions, and emission-line
fluxes are instead included in the fit, the best-fit black hole mass
decreased by 11\% to $3.1 \times 10^9\ M_\odot$.

\smallskip

\noindent \emph{Asymmetric Drift Correction}: By assigning a dynamical
origin to the intrinsic velocity dispersion and calculating an
asymmetric drift correction as discussed in \S
\ref{subsec:asymmdrift}, the black hole mass increased by 6\% to $3.7
\times 10^9\ M_\odot$. Such a minor change to $M_\mathrm{BH}$ is not
surprising considering the small amount of intrinsic velocity
dispersion relative to the large velocities that are seen in
M87. Compared to the dynamically cold, thin-disk model, the model with
an asymmetric drift correction was a worse fit to the observed
velocities, with a $\chi^2 = 71$.

\begin{figure*}
\begin{center}
\epsscale{1.0}
\plotone{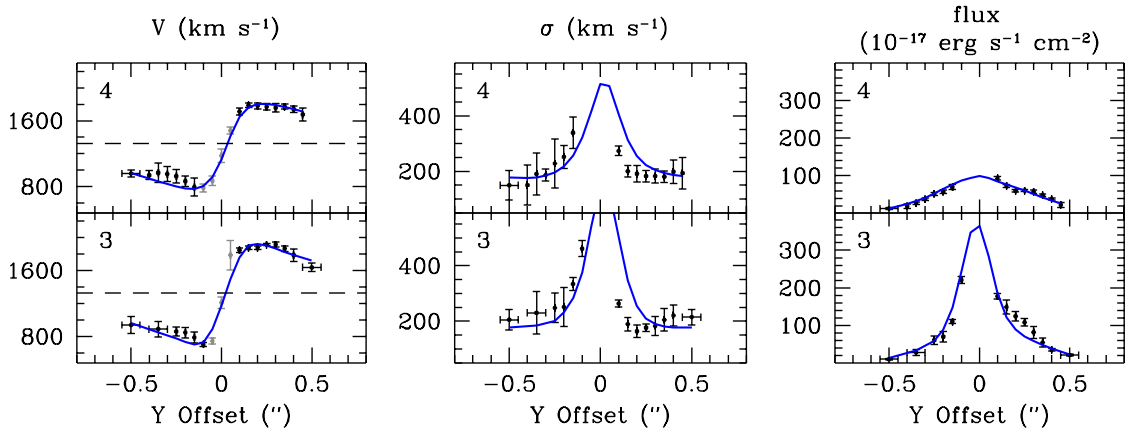}
\caption{The predictions of a disk model run with an asymmetric drift
  correction are compared to the data from slit positions 3 and 4; see
  Figure \ref{fig:datamodel} for additional description. Here, an
  extreme intrinsic velocity dispersion has been chosen so that the
  model would match the large observed line widths located
  $-$0\farcs10 and $-$0\farcs15 away from the center of slits 3 and 4,
  at the expense of being able to reproduce the smaller observed line
  widths at some of the other locations. This exercise provides an
  indication of the maximum possible increase in $M_\mathrm{BH}$ when
  assigning a dynamical origin to the intrinsic velocity
  dispersion. \label{fig:maxasymmdrift}}
\end{center}
\end{figure*}

The predictions from the model with asymmetric drift correction,
however, continue to be significantly smaller (by $\sim$140 $-$ 240 km
s$^{-1}$) than the observed dispersions located $-$0\farcs10 and
$-$0\farcs15 away from the center of slits 3 and 4. In order to
establish the maximum possible increase in the black hole mass, we
searched for a new parameterization of the intrinsic velocity
dispersion that would match these few points, at the expense of being
able to reproduce the observed velocity dispersions at the other
locations. Since the distribution of the observed line widths is
asymmetric, and the velocity dispersions at negative Y Offsets are
generally larger than those measured at positive Y Offsets, we
modified the input kinematics by simply replacing the observed
velocity dispersions from the positive Y Offset side with the values
from the negative Y Offset side. We then fit gas-dynamical models with
an asymmetric drift correction first to the modified line widths and
then to the velocities. We determined that the intrinsic velocity
dispersion was best characterized as a constant $+$ exponential
function, with $\sigma_0 = 202$ km s$^{-1}$, $\sigma_1 = 2040$ km
s$^{-1}$, and $r_0 = 4.4$ pc, and that the black hole mass increased
by 20\% to $M_\mathrm{BH} = 4.2 \times 10^9\ M_\odot$. As can be seen
in Figure \ref{fig:maxasymmdrift}, the new model is able to
sufficiently match the high velocity dispersions observed at Y Offsets
of $-$0\farcs10 and $-$0\farcs15 from slits 3 and 4.

\smallskip

The formal model fitting error and the additional sources of
uncertainty were added in quadrature to derive the final range of
black hole masses. Thus, we find the M87 black hole mass is (2.8 $-$
4.4) $\times 10^9\ M_\odot$ (1$\sigma$ uncertainties) and (2.7 $-$
6.9) $\times 10^9\ M_\odot$ (3$\sigma$ uncertainties) with a best-fit
value of $3.5 \times 10^9\ M_\odot$. We propagated the 6\% increase in
the black hole mass resulting from the asymmetric drift correction
into the upper uncertainty of $M_\mathrm{BH}$ only. Even when using an
extreme characterization of the intrinsic velocity dispersion to
better match the observed line widths at Y Offsets of $-$0\farcs10 and
$-$0\farcs15 in slits 3 and 4, the black hole mass does not increase
beyond the final 1$\sigma$ uncertainties.

\section{Discussion and Conclusions}
\label{sec:discussion_conclusions}

Using newly acquired STIS observations of M87 with the 0\farcs1-wide
aperture at five adjacent parallel positions, we have measured the
[\ion{N}{2}] velocity, velocity dispersion, and flux well within the
black hole's dynamical sphere of influence. Through gas-dynamical
modeling of the STIS data, we have determined that the mass of the
black hole is $M_\mathrm{BH} = (3.5_{-0.7}^{+0.9}) \times 10^9\
M_\odot$. This black hole mass is similar to the expectations from the
$M_\mathrm{BH}$ $-$ bulge relations. With $\sigma_\star = 324$ km
s$^{-1}$ \citep{Gebhardt_2011} and a $V$-band luminosity calculated
from $M_V = -22.71$ \citep{Lauer_2007}, the most recent calibrations
of the $M_\mathrm{BH} - \sigma_\star$ relation suggest that
$M_\mathrm{BH}$ $\sim$ $3 \times 10^9\ M_\odot$
\citep{Graham_Scott_2013, McConnell_Ma_2013} and the $M_\mathrm{BH} -
L_\mathrm{bul}$ correlation predicts $M_\mathrm{BH} = 2 \times 10^9\
M_\odot$ \citep{McConnell_Ma_2013}.

Unlike the past gas-dynamical studies of \cite{Harms_1994} and
\cite{Macchetto_1997}, we have mapped out the complete kinematic
structure of the emission-line disk. We have constrained the gas disk
inclination angle, which has previously been a source of uncertainty,
to be between $35^\circ < i < 47^\circ$ with a best-fit value of $i =
42^\circ$. In addition, our models included a detailed treatment of
the effects of the telescope and instrument. For the first time, we
have found that a relatively small amount of additional velocity
dispersion internal to the gas disk is needed to reproduced the
observed line widths. \cite{Macchetto_1997} did not find evidence for
such an intrinsic velocity dispersion in the M87 disk, but the
instrumental resolution of their FOC observations was low (FWHM $\sim$
430 km s$^{-1}$). Thus, the order-of-magnitude improvement in spectral
resolution of STIS compared to the older FOC data has given us the
sensitivity to detect an intrinsic velocity dispersion. We considered
the possibility that the intrinsic velocity dispersion provides
dynamical support to the gas disk, and determined that inferred mass
increases by just 6\%. We incorporated this effect, along with a
number of other sources of uncertainty, into the error budget.

Before comparing our mass measurement to the prior mass
determinations, there are some additional systematics to keep in mind
when interpreting the results. Our thin-disk model matches the overall
shape of the observed velocity curves very well and returns
$\chi^2_\nu = 1.2$, however the data is not randomly scattered about
the best-fit model at all locations. Instead the model systematically
deviates from the observations in certain regions, at Y Offset =
0\farcs25 $-$ 0\farcs4 at slit 3, for example. Consequently, there may
be some regions that depart from circular rotation, whose small-scale
velocity structure cannot be matched with any disk model. This is a
common problem for gas-dynamical models in general, and makes
estimating the uncertainties in the black hole mass challenging. Past
studies have opted to rescale the error bars on the kinematics, such
that the best-fit model would have a $\chi^2_\nu \approx 1$, prior to
measuring the statistical errors on $M_\mathrm{BH}$
\citep[e.g.,][]{Barth_2001, Atkinson_2005, deFrancesco_2006,
  DallaBonta_2009}. For this work, we do not adjust the kinematic
errors because our best-fit model already has $\chi^2_\nu \approx
1$. Despite the reasonable $\chi^2_\nu$ of our best-fit model, we note
that the model is not a perfect description of the gas kinematics
everywhere in the disk.

Furthermore, we treat the (small amount of ) intrinsic velocity as
dynamically important by applying an asymmetric drift correction. The
correction is applicable in the limit of collisionless particles and
for $\sigma_r/v_c \ll 1$. Here, we applied the correction to gas
clouds, which are not collisionless, and to a system that has
$\sigma_r/v_c < 0.3$ over the radial extent of the STIS
kinematics. Also, the asymmetric drift correction depends upon the
number density of clouds in the disk, $\nu(r)$, which is
unknown. While we assumed that the intrinsic emission-line flux
distribution can be used as a proxy for $\nu(r)$, this is at best a
rough approximation. With these considerations in mind, our model with
an asymmetric drift correction only provides estimation, albeit a
useful one, of the dynamical influence of the intrinsic velocity
dispersion.

Our mass measurement is consistent with the two previous gas-dynamical
measurements of the black hole in M87. The \cite{Harms_1994} and
\cite{Macchetto_1997} masses are $M_\mathrm{BH} = (2.9 \pm 0.8) \times
10^9\ M_\odot$ and $M_\mathrm{BH} = (3.8 \pm 1.1) \times 10^9\
M_\odot$, after scaling to our adopted distance. Of the remaining disk
model parameters, our best-fit inclination angle is the same as the
one used by \cite{Harms_1994} and is consistent with the lower bound
of acceptable values ($i = 47^\circ - 65^\circ$) found by
\cite{Macchetto_1997}, while our best-fit systemic velocity
($\upsilon_\mathrm{sys} = 1335$ km s$^{-1}$) is larger than the ones
used by \cite{Harms_1994} ($\upsilon_\mathrm{sys} = 1309$ km s$^{-1}$)
and determined by \cite{Macchetto_1997} ($\upsilon_\mathrm{sys} =
1290$ km s$^{-1}$). For comparison, the recession velocities measured
from optical lines given by the NASA Extragalactic Database span a
wide range of values, from $800 - 1747$ km s$^{-1}$. We established
that the projected major axis of the gas disk is $45^\circ$ east of
north, which is roughly perpendicular to the jet and similar to
position angle of $38^\circ$ measured by \cite{Macchetto_1997}.

The most recent M87 stellar-dynamical black hole measurement of
$M_\mathrm{BH} = (6.6 \pm 0.4) \times 10^9\ M_\odot$
\citep{Gebhardt_2011} is a factor of two larger than our mass
measurement, and there is a 2$\sigma$ discrepancy between our
gas-dynamical mass and the \cite{Gebhardt_2011} stellar-dynamical mass
when accounting for the uncertainties associated with each
measurement. Recent stellar-dynamical work has shown the importance in
some galaxies of including a dark halo, incorporating orbital
libraries that better sample the phase-space occupied by tangential
orbits, using triaxial geometries, and accounting for a spatially
varying stellar mass-to-light ratio \citep{Gebhardt_Thomas_2009,
  Shen_Gebhardt_2010, vandenBosch_deZeeuw_2010, Schulze_Gebhardt_2011,
  McConnell_2013}. While \cite{Gebhardt_2011} address the first two
potential sources of systematic error, they do not examine the later
two sources. In particular, they do not fit for the galaxy's intrinsic
shape and orientation, and instead use axisymmetric orbit-based models
that assume an edge-on inclination. Therefore, it is still possible
for the stellar-dynamical black hole mass measurement to change again
if more general triaxial models are applied, although the magnitude
and direction of the effect on the inferred black hole mass is unknown
a priori. Also, \cite{Gebhardt_2011} construct stellar-dynamical
models under the assumption of a constant stellar mass-to-light ratio,
however, as shown by \cite{McConnell_2013}, spatial gradients in the
mass-to-light ratio can have a non-negligible effect on
$M_\mathrm{BH}$. In contrast, our dynamical models rely on gas in
circular orbits at small radii and are insensitive to the overall
normalization of the stellar mass-to-light ratio (\S
\ref{subsec:mbherror}), making our gas-dynamical measurement immune to
this uncertainty.

Our gas-dynamical measurement and the stellar-dynamical determination
of \cite{Gebhardt_2011} utilize high-quality, high-resolution
observations and apply some of the most up-to-date modeling
techniques, yet there continues to be a discrepancy in the M87 black
hole mass. This highlights the need for carrying out more such
cross-checks between the two methods. Currently, M87 is one of six
consistency tests that have yielded a meaningful comparison between
the gas and stellar-dynamical techniques, and in three cases (this
study included) the stellar-dynamical mass exceeds the gas-dynamical
determination by a factor of $\sim$2 $-$ 5. Since the stellar and
gas-dynamical techniques are completely independent methods with
different systematic effects, only by applying both techniques to the
same object can conclusions be made regarding the consistency of the
methods, the subsequent effects on the $M_\mathrm{BH}$ scaling
relations, and the magnitude and distribution of the intrinsic scatter
of the $M_\mathrm{BH}$ $-$ host galaxy relations.

\acknowledgements

J.~L.~W. has been supported by an NSF Astronomy and Astrophysics
Postdoctoral Fellowship under Award No. 1102845. Support for program
\#12162 was provided by NASA through a grant from the Space Telescope
Science Institute, which is operated by the Association of
Universities for Research in Astronomy, Inc., under NASA contract NAS
5-26555. Research by A.~J.~B. has additionally been supported by NSF
grant AST-1108835. Some of the data presented in this paper were
obtained from the Multimission Archive at the Space Telescope Science
Institute (MAST). STScI is operated by the Association of Universities
for Research in Astronomy, Inc., under NASA contract NAS 5-26555. This
research has made use of the NASA/IPAC Extragalactic Database (NED)
which is operated by the Jet Propulsion Laboratory, California
Institute of Technology, under contract with NASA. We dedicate this
paper to the memory of Wallace L.~W. Sargent, whose early work on M87
has had a profound impact on the field of black hole
searches. A.~J.~B. and L.~C.~H. have also benefited enormously from
many years of interaction and fruitful scientific collaboration with
Wal.

\end{document}